\documentstyle[11pt]{article}
\newcommand{\beq}{\begin{equation}}
\newcommand{\eeq}{\end{equation}}

\newcommand{\beqs}{\begin{eqnarray}}
\newcommand{\eeqs}{\end{eqnarray}}

\newcommand{\nn}{\nonumber}

\begin{document}

%\hfill HEP-TH
\begin{center}
\vskip 2.5cm
{\LARGE \bf New Prescription in light-cone gauge theories}

\vskip 1.0cm
{\Large  D.~K.~Park }
\\
{\large  Department of Physics,  KyungNam University, Masan, 631-701
Korea}

\vskip 0.4cm
\end{center}

\centerline{\bf Abstract}

New prescription for the singularities of light-cone gauge theories
is suggested.
The new prescription provides Green's function which is identical
with and different from that of Mandelstam-Leibbrandt prescription
at $d=4$ and $d=2$ respectively.
\vfill

\newpage
\setcounter{footnote}{1}

\newcommand{\tr}{\;{\rm tr}\;}
The light-cone gauge(radiation gauge in light-cone coordinate), one of the
latest noncovariant gauge, has been
frequently used for the calculation of perturbative QCD, the quantization
of the supersymmetric Yang-Mills theories, and the non-covariant
formulation of string theories in spite of the lack of manifest
Lorentz covariance.[1-4]

However, implementation of light-cone gauge is not entirely straightforward,
at least in perturbation theory. This statement is easily described by
considering the free gauge field propagator[5]
\beq
G^{ab}_{\mu \nu}(k) =
      \frac{-i \delta^{ab}}{k^2}
      \left[ g_{\mu \nu} - \frac{n_{\mu}k_{\nu} + n_{\nu}k_{\mu}}{n \cdot k}
                                                            \right].
\eeq
 From Eq.(1) one can see that there are two kinds of singularities in
$G^{ab}_{\mu \nu}(k)$. First singularity arises when $k^2 = 0$.
This is the univeral property of massless fields.
Usually this singularity is prescribed by choosing a causal
prescription,
\beq
 \frac{1}{k^2} \rightarrow \frac{1}{k^2 + i \epsilon}.
\eeq
In $G^{ab}_{\mu \nu}(k)$ there is another singularity which arises
when $n \cdot k = 0$. This "spurious singularity" is peculiar one
in light-cone gauge.[6]

For the last decade various prescriptions have been made for the
spurious singularity.
If a Cauchy principal
value(CPV) prescription
\beq
   \frac{1}{k^-} \rightarrow CPV(\frac{1}{k^-}) \equiv
                               \frac{1}{2} \left[
                               \frac{1}{k^-+i\epsilon} +
                               \frac{1}{k^--i\epsilon}
                               \right],
\eeq
which palys an crucial role in other non-covariant gauges is choosed,
calculation of the various Feynman diagrams produces a poorly defined
integrals.[7] A more successful prescription for the spurious singularity
which is usually called by Mandelstam-Leibbrandt(ML) prescription
\beq
   \frac{1}{k^-} \rightarrow ML(\frac{1}{k^-}) \equiv
                                \frac{k^+}{k^+ k^- + i \epsilon}
\eeq
is suggested independently by Mandelstam[8] and Leibbrandt[9].
Later it is proved in the framework of equal-time canonical quantization
that ML prescription is nothing but the causal prescription[10] and
the renormalizibility of the gauge theories formulated in this way is
also shown[11] although nonlocal counterterms are necessary to render
off-shell Green's function finite.

Although attention is paid only for the spurious singularity for the
last decade, there was a suggestion for $k^2 = 0$ singularity on
two dimensional light-cone about twenty years ago.[12] The authors
in Ref.[12] suggested that the propagators on the two-dimensional
light-cone is different from those in conventional coordinates as follows
by analyzing the massless scalar and fermion theories
\beq
    \frac{1}{k^2 + i \epsilon} \Rightarrow
         \frac{1}{k^2 + i \epsilon} +
              \frac{i \pi}{2} \frac{\delta(k^-)}{\mid k^+ \mid}.
\eeq
Recently it is shown [13] that the difference of the propagator
on the two-dimensional light-cone also can be interpreted as the
difference of prescription like
\beq
    \frac{1}{2k^+} ML(\frac{1}{k^-}) \Rightarrow
    \frac{1}{2k^+} CPV(\frac{1}{k^-}).
\eeq
This means that the prescription problem arises not only
in the spurious singularity but also in $k^2 = 0$ singularity
in the light-cone gauge theories. In this paper, therefore, we will
choose the prescriptions for $k^2 = 0$ and spurious singularities
simultaneously.

In order to find the new prescription let us consider only
$G^{ab}_{--}(k)$ which is the only non-vanishing component in
two-dimensional theory. If one chooses ML-prescription
\beq
[G^{ab}_{--}(k)]_{ML} =
  \frac{2i\delta^{ab}k^+}{k^2 + i\epsilon}
  ML(\frac{1}{k^-}),
\eeq
then $(-,-)$component of d-dimensional Green's function defined as
\beq
D^{ab}_{\mu \nu}(x) \equiv
                    \frac{1}{(2 \pi)^d}
                    \int d^dk G^{ab}_{\mu \nu}(k) e^{ikx}
                    \equiv
                    \delta^{ab} D_{\mu \nu}(x)
\eeq
is
\beqs
\lefteqn{[D_{--}(x)]_{ML}}  \\ \nn
&=& \frac{\Gamma(\frac{d}{2})}{2 \pi^{\frac{d}{2}}}
                          (-x^2 + i \epsilon)^{-\frac{d}{2}} (x^+)^2 \\ \nn
& &      \times           \left[ 2 _2F_1(1, \frac{d}{2} - 1; 2;
                          \frac{2 (n \cdot x) (n^{\ast} \cdot x)}{(n \cdot
                          n^{\ast}) x^2})
                          + (\frac{{\bf x_T}^2}{x^2}) _2F_1(2, \frac{d}{2}; 3;
                          \frac{2 (n \cdot x) (n^{\ast} \cdot x)}{(n \cdot
                          n^{\ast}) x^2})
                    \right]
\eeqs
where $_2F_1(a, b; c; z)$ is usual hypergeometric function.
One can show that $ d \rightarrow 4$ limit of Eq.(9) coincides
with the result of Ref.[14] if the difference of definition of
$n^{\mu}$ is considered. If one takes ${\bf x_T} \rightarrow {\bf 0}$
limit, Eq.(9) becomes
\beq
\lim_{{\bf x_T} \rightarrow {\bf 0}} [D_{--}(x)]_{ML} =
\frac{2 \Gamma(\frac{d}{2})(x^+)^2}{\pi^{\frac{d}{2}}(4 - d)}
(-x^2 + i \epsilon)^{-\frac{d}{2}}
\eeq
whose $ d \rightarrow 2 $ limit is
\beq
[D_{--}(x)]^{d=2}_{ML} = \frac{(x^+)^2}{\pi}
                         \frac{1}{-x^2 + i\epsilon}.
\eeq
This is different from
\beq
[D_{--}(x)]_{'t Hooft} = - \frac{i}{2} \mid x^+ \mid \delta(x^-)
\eeq
which was used by 't Hooft in Ref.[15] for the calculation of the
mesonic mass spectrum. This difference makes the authors of Ref.[16] suggest
that the two-dimensional pure Yang-Mills theory with light-cone
gauge is not free theory. Their suggestion arises from the fact
that calculational result of the vacuum expectation value of the
lightlike Wilson-loop operator with ML-prescription at $O(g^4)$
does not exhibit abelian exponentiation. So it is worthwhile
to check whether there exists a prescription which provides
a Green's function whose $d \rightarrow 2$ limit coincides with
$ [D_{--}(x)]_{'t Hooft} $. Soon it will be shown that this will
be achieved by choosing CPV-prescription for $k^2 = 0$ and spurious
singularities simultaneously like
\beqs
\lefteqn{G^{ab}_{--}(k)} \\ \nn
&=& \frac{2i\delta^{ab}k^+}{k^2} \frac{1}{k^-} \\ \nn
&\rightarrow&
[G^{ab}_{--}(k)]_{NCPV}  \\ \nn
&\equiv& i \delta^{ab} CPV \left(
                                           \frac{1}{k^-(k^- -
                                            \frac{{\bf k_T}^2}{2k^+})}
                                    \right)  \\ \nn
&=& i \delta^{ab} \frac{2k^+}{{\bf k_T}^2}
                  \left[ CPV(\frac{1}{k^- - \frac{{\bf k_T}^2}{2k^+}})
                         - CPV(\frac{1}{k^-})
                  \right].
\eeqs
By using the formula
\beqs
\lefteqn{CPV(\frac{1}{k^- - \frac{{\bf k_T}^2}{2k^+}})} \\ \nn
&=& \frac{1}{k^- - \frac{{\bf k_T}^2}{2k^+} + i \epsilon \epsilon(k^+)}
  + i \pi \epsilon(k^+) \delta( k^- - \frac{{\bf k_T}^2}{2k^+}), \\ \nn
\lefteqn{CPV(\frac{1}{k^-})} \\ \nn
&=& ML(\frac{1}{k^-}) + i \pi \epsilon(k^+)
                     \delta(k^-),
\eeqs
$[G^{ab}_{--}(k)]_{NCPV}$ becomes
\beq
[G^{ab}_{--}(k)]_{NCPV} = [G^{ab}_{--}(k)]_{ML} - 2 \pi \delta^{ab}
                          \frac{\mid k^+ \mid}{{\bf k_T}^2}
                          \left[
                                \delta(k^- - \frac{{\bf k_T}^2}{2 k^+})
                                - \delta(k^-)
                          \right].
\eeq
 From Eq.(15) $[D_{--}(x)]_{NCPV}$ is directly calculated and the final
result is
\beq
[D_{--}(x)]_{NCPV} = [D_{--}(x)]_{ML} + \Delta D_{--}(x)
\eeq
where $[D_{--}(x)]_{ML}$ is given in Eq.(9) and $\Delta D_{--}(x)$ is
\beqs
\lefteqn{\Delta D_{--}(x)} \\ \nn
&=&  - 2^{2 - d} \pi^{- \frac{d}{2}}
                   \sum_{l = 1}^{\infty} \frac{\Gamma(l + \frac{d}{2} - 2)}
                                              {l! \Gamma(1 - l)}
                   (- \frac{x^+}{2})^l (\frac{{\bf x_T}^2}{4})^{2 - \frac{d}{2}
                                                                - l}
                   (\partial_-)^{-l+1} CPV(\frac{1}{x^-}).
\eeqs
The modification term $\Delta D_{--}(x)$ does not give a finite contribution
at $ d = 4 $. Therefore this new prescription provides a same
$D_{--}(x)$ with ML-prescription. After the calculation of other
components one can show that all components of $[D_{\mu \nu}(x)]_{NCPV}$
coincide with $[D_{\mu \nu}(x)]_{ML}$ at $d = 4$.

However the situation is completely different at $d = 2$. In this case $\Delta
D_{--}(x)$
gives a finite contribution when $ l = 1$. By considering this finite
contribution $[D_{--}(x)]_{NCPV}$ at $d = 2$ becomes
\beqs
\lim_{d \rightarrow 2}[D_{--}(x)]_{NCPV}
&=& \frac{(x^+)^2}{\pi} \frac{1}{-x^2 + i \epsilon}
    + \frac{x^+}{2 \pi} CPV(\frac{1}{x^-})          \\ \nn
&=& - \frac{i}{2} \mid x^+ \mid \delta(x^-),
\eeqs
which is exactly same with $[D_{--}(x)]_{'t Hooft}$.
Therefore this new prescription provides a same Green's function with
ML-prescription at $d = 4$ and 't Hooft approach at $d = 2$.
Furthermore if one follows this new prescription, one can not say that two
dimensional pure Yang-Mills theory with light-cone gauge is interacting
theory which is suggested in Ref.[16]. For example let us calculate
the crossed diagram of lightlike Wilson-loop operator which gives
a non-vanishing and vanishing contributions if one chooses a ML-prescription
and 't Hooft approach respectively. After following the notation of
Ref.[16] this new CPV prescription gives
\beqs
\lefteqn{[W_{crossed}]_{CPV}} \\ \nn
&=& - \frac{1}{2} (ig)^4 \mu^{4 - 2 d}  C_F C_A
    (n^{\ast -})^4
    \int_{0}^{1} ds_1 \int_{0}^{s_1} ds_2
    \int_{1}^{0} dt_1 \\ \nn
& & \int_{1}^{t_1} dt_2
   [D_{--}(n + n^{\ast}(t_1 - s_1))]_{NCPV}
   [D_{--}(n + n^{\ast}(t_2 - s_2))]_{NCPV} \\ \nn
&=& [W_{crossed}]_{ML} + \Delta W_{crossed}.
\eeqs
$[W_{crossed}]_{ML}$ is already calculated in Ref.[16];
\beqs
\lefteqn{[W_{crossed}]_{ML}} \\ \nn
&=& -(\frac{g}{\pi \mu})^4 \frac{C_F C_A}{16}
                    \frac{\Gamma^2(\frac{d}{2} - 1)}{(d - 4)^2}
                    \left[ 2 A \frac{d - 2}{d - 3} +
                           8 B \left( 1 - 2 \frac{\Gamma^2(3 -
                                     \frac{d}{2})}{\Gamma(5 - d)}
                               \right)
                     \right],
\eeqs
where
\beqs
      A&=& (2 \pi \mu^2 n \cdot n^{\ast} + i \epsilon)^{4 - d} +
          (-2 \pi \mu^2 n \cdot n^{\ast} + i \epsilon)^{4 - d}, \\ \nn
      B&=& [(2 \pi \mu^2 n \cdot n^{\ast} + i \epsilon)
           (-2 \pi \mu^2 n \cdot n^{\ast} + i \epsilon)]^{2 - \frac{d}{2}},
\eeqs
and this gives a finite contribution at $d=2$
\beq
\lim_{d \rightarrow 2} [W_{crossed}]_{ML} = \frac{g^4}{48}
                 C_F C_A (n \cdot n^{\ast})^2.
\eeq
In order to calculate $\Delta W_{crossed}$ we divide it as two parts
\beq
        \Delta W_{crossed} = \Delta W_1 + \Delta W_2
\eeq
where
\beqs
\lefteqn{\Delta W_1} \\ \nn
&=&   -\frac{1}{2} (ig)^4 \mu^{4 - 2d} C_F C_A (n^{\ast -})^4
                \int_0^1 ds_1 \int_0^{s_1} ds_2
                \int_0^1 dt_1 \int_{t_1}^1 dt_2 \\ \nn
& &\bigg[ [D_{--}(n + n^{\ast}(t_1 - s_1))]_{ML}
           \Delta D_{--}(n + n^{\ast}(t_2 - s_2))             \\ \nn
& &            +
                    [D_{--}(n + n^{\ast}(t_2 - s_2))]_{ML}
           \Delta D_{--}(n + n^{\ast}(t_1 - s_1))
                \bigg]
\eeqs
and
\beqs
\lefteqn{\Delta W_2} \\ \nn
&=& -\frac{1}{2} (ig)^4 \mu^{4 - 2d} C_F C_A (n^{\ast -})^4
                 \int_0^1 ds_1 \int_0^{s_1} ds_2
                 \int_0^1 dt_1 \int_{t_1}^1 dt_2 \\ \nn
& &  \Delta D_{--}(n + n^{\ast}(t_1 - s_1)) \Delta D_{--}(n + n^{\ast}(t_2 -
s_2)).
\eeqs
After tedious calculation one can show that $\Delta W_1$ and $\Delta W_2$
provide finite contribution to
$[W_{crossed}]_{NCPV}$ at $d = 2$
\beqs
\lim_{d \rightarrow 2} \Delta W_1 = - \frac{g^4}{24}
                 C_F C_A (n n^{\ast})^2 \\ \nn
  \lim_{d \rightarrow 2} \Delta W_2 = \frac{g^4}{48}
                 C_F C_A (n n^{\ast})^2 \\ \nn
\eeqs
from which the vanishing of $[W_{crossed}]_{NCPV}$ can be proved.
This result is in agreement with that of 't Hooft approach and differs
from that of ML-prescription. So it is worthwhile to calculate the
remaining $O(g^4)$ diagrams(self-energy and vertex diagrams) to check
whether the two-dimensional Yang-Mills theory with light-cone gauge
is free or not by using this new prescription. This work will be
reported elsewhere.

\bigskip

\noindent {\em ACKNOWLEDGMENTS}

This work was carried out with support from the Korean Science and
Engineering Foundation.

\end{document}